# SeisGPT: A Physics-Informed Data-Driven Large Model for Real-Time Seismic Response Prediction

**Shiqiao Meng[1], Ying Zhou[1]*, Qinghua Zheng[2], Bingxu Liao[1], Mushi Chang[1], Tianshu Zhang[1], Abderrahim Djerrad[1]**

[1]State Key Laboratory of Disaster Reduction in Civil Engineering, Tongji University, Shanghai, China
[2]School of Computer Science and Technology, Tongji University, Shanghai, China

**Abstract**

Accurately predicting the dynamic responses of building structures under seismic loads is essential for ensuring structural safety and minimizing potential damage. This critical aspect of structural analysis allows engineers to evaluate how structures perform under various loading conditions, facilitating informed design and safety decisions. Traditional methods, which rely on complex finite element models often struggle with balancing computational efficiency and accuracy. To address this challenge, we introduce SeisGPT, a data-driven, large physics-informed model that leverages deep neural networks based on the Generative Pre-trained Transformer (GPT) architecture. SeisGPT is designed to predict, in real-time the dynamic behavior of building structures under seismic forces. Trained on a diverse corpus of seismic data and structural engineering principles, it instantly generates predictive responses, including displacement, acceleration, and inter-story drift, with high accuracy and computational efficiency. Its adaptability across various building typologies and seismic intensities makes this framework a valuable tool for designing robust structures and assessing seismic risk. Through comprehensive validation, this approach exhibits superior performance, offering engineers and researchers a powerful tool for assessing seismic response and informing resilient design strategies. This innovative framework represents a significant advancement in seismic engineering practice, with potential applications in mitigating seismic hazards and enhancing structural resilience.

## 1. Introduction

Earthquakes rank among the most devastating natural disasters, causing significant loss of life and property damage. The collapse and failure of buildings during major seismic events are primary contributors to these losses [1]. While predicting the exact timing and location of earthquakes remains challenging, structural engineering aims to accurately forecast how buildings will respond to seismic forces [2]. Precise and efficient prediction of structural responses are crucial for disaster prevention and mitigation, informing key tasks such as post-earthquake damage assessment, structural health monitoring, and resilience evaluation [3, 4]. Efforts to improve seismic response prediction have led to various approaches, including empirical methods and simplified analytical models [5, 6]. However, these methods often oversimplify complex behaviors, lack generality, and struggle to account for diverse building typologies. High-precision structural dynamic analysis heavily relies on the finite element method (FEM) [7, 8], a widely used simulation technique. Nevertheless, even with advancements in computational power, the growing complexity of numerical models leads to substantial computational demands, particularly for problems exhibiting nonlinear hysteretic behavior under dynamic loads. This situation presents significant challenges in achieving a balance between computational efficiency and accuracy.

The rapid advancement of machine learning and deep learning technologies has ushered in a new era in structural engineering [9, 10], particularly in the domain of seismic response prediction and approximation for buildings [11, 12]. Researchers in this field are actively exploring and developing innovative strategies and models that leverage these cutting-edge computational techniques. These approaches can be broadly categorized into two main streams: data-driven methods [13, 14] and physics-informed approaches [15-17]. Data-driven methods rely heavily on historical data and observed patterns to make predictions. They attempt to uncover complex patterns and relationships in structural responses without explicitly modeling the underlying physical processes. On the other hand, physics-informed approaches aim to incorporate domain knowledge and fundamental principles of structural dynamics into the machine learning framework. These methods seek to enhance the reliability and generalizability while potentially reducing the amount of data required for training.

Recent advancements in deep learning have significantly impacted this field. Algorithms such as Convolutional Neural Networks (CNNs) [18-20], Recurrent Neural Networks (RNNs) [21-24], Long Short-Term Memory (LSTM) networks [25-27], and Transformers have revolutionized seismic response prediction. These powerful models excel in pattern recognition and data fitting, especially for time series problems. Their ability to learn intricate patterns from extensive datasets enables more precise and detailed predictions. Consequently, they can extract complex temporal relationships from seismic data, resulting in highly accurate and nuanced forecasts of structural behavior during earthquakes. This exciting development marks a significant shift from traditional, purely physics-based methods, offering a new paradigm in understanding and forecasting the seismic behavior of various building types. The application of machine learning to predict the behavior of structures under dynamic loads has been explored by several researchers, including Zhang [25], and Eshkevari [22]. Their studies highlight the significant potential of machine learning in advancing structural dynamics and response predictions. Zhang [28] in particular, employed physics-based LSTM models for predicting seismic responses, demonstrating the effectiveness of these approaches in capturing structural behavior under seismic loads.

Despite significant advancements, several key challenges remain in the field [29]. Many current approaches rely on overly simplified machine learning models, which hinder their ability to capture the complexities of real-world scenarios. A major limitation is the difficulty in achieving both high accuracy and detailed response predictions across numerous floors within intricate structures [25]. Existing methods often focus on simpler buildings, neglecting the unique behaviors and intricacies of high-rise, large and complex structures. Limitations in data availability and a prioritization of computational speed contribute to this oversight, along with the exclusion of critical structural physics information. Consequently, the precision of these predictions is frequently inadequate, leading to suboptimal outcomes, which emphasizes the need for advanced models capable of predicting specific failure modes in complex structural elements [30]. These challenges highlight the necessity for comprehensive models that can accurately capture the dynamic behavior of a wide variety of building types under different seismic intensities.

To address the aforementioned challenges, we introduce SeisGPT, a novel large model that leverages deep neural networks based on the Generative Pre-trained Transformer (GPT) architecture. This cutting-edge, data-driven, and physics-informed large model represents a significant leap forward in structural engineering, offering real-time prediction of dynamic responses in complex building structures under seismic forces with unprecedented accuracy and computational efficiency. SeisGPT aims to address the shortcomings of traditional machine learning models by integrating the strengths of data-driven learning with physics-informed constraints. This innovative deep learning framework incorporates critical structural dynamic features, which simplifies the model fitting process compared to methods that rely solely on data. As a result, SeisGPT achieves enhanced prediction accuracy across a wide range of building types and seismic intensities, effectively striking a balance between precision and computational efficiency.

Our large model brings several key innovations to the field. It integrates crucial structural physical information, especially dynamic characteristics, enabling the capture of complex relationships between structural properties, seismic excitations, and resulting structural responses. The model provides real-time prediction capabilities for different floors and directions within a building, facilitating comprehensive damage assessment. Extensive experiments using real-world data demonstrate the significant performance advantages of our large model over competitive baselines. SeisGPT excels in prediction precision, computational efficiency, and applicability across various building typologies and seismic intensities.

Rigorous validation further confirms the framework's robustness and reliability in accurately predicting structural responses under seismic actions.

The potential applications of SeisGPT, as a large model for structural engineering, extend far beyond immediate seismic response prediction. Its capabilities can revolutionize post-earthquake damage assessment, structural health monitoring, and seismic resilience evaluation. Furthermore, the framework is well-suited for computationally intensive analyses such as fragility and reliability analysis, nonlinear dynamic analysis, performance-based earthquake engineering (PBEE) [31], probabilistic seismic hazard analysis (PSHA) [32], and incremental dynamic analysis (IDA) [33, 34]. By offering a more accurate and efficient solution for predicting structural responses in complex building structures, SeisGPT, provides engineers and researchers with a valuable tool for designing resilient structures and assessing seismic risk. Its broad applicability and advanced capabilities make it an essential asset in the development of comprehensive seismic design strategies and detailed performance assessments. The introduction of SeisGPT marks a significant step forward in our ability to understand, predict, and mitigate the effects of seismic events on built environments, potentially saving lives and reducing economic losses in earthquake-prone regions worldwide.

# 2. Methodology

## 2.1. SeisGPT Workflow Overview

The SeisGPT framework utilizes a large, deep learning-based model to deliver precise and computationally efficient solution for real-time predictions of building structure responses under seismic loads. By eliminating the need for complex finite element (FE) modeling, SeisGPT significantly reduces the computational burden typically associated with traditional FE methods, resulting in substantial savings in both time and energy resources. A schematic of the framework is provided in Figure 1.

The SeisGPT framework employs a two-phase deep learning strategy. Initially, a pre-trained model, termed "SeisGPT-Base," is trained on an extensive dataset encompassing responses of various building typologies and seismic intensities. This training process equips the model with a foundational understanding of how structures respond to seismic loads. To further enhance accuracy for specific cases and scenarios, this pre-trained base model can be fine-tuned using task-specific data, resulting in the "SeisGPT-Enhanced" model. This fine-tuning process tailors the model capabilities to particular building types or regions, potentially leading to superior prediction accuracy for those specific applications.

The development of the SeisGPT framework consists of several key components, each crucial for building robust pre-trained and fine-tuned deep learning models:

1. The first component is the *Numerical Module*, a crucial component in the framework that mainly serves as a data generator. A vast collection of representative seismic waves is carefully selected to simulate various earthquake scenarios and capture different ground motion characteristics. Simultaneously, detailed macro models representing diverse building structure typologies are created. These selected seismic waves are used in conjunction with macro models to perform Nonlinear time history analysis (NLTHA), generating a large dataset of structural response time histories, including displacements, accelerations, and other relevant parameters. The dataset, covering a wide range of seismic events and building types, forms the foundation for model training.

2. Following, the *Simplified Dynamic Response (SDR) Module* is designed to extract key structural dynamic features—such as story mass and stiffness—from a detailed macro model. These features are then used to create a simplified lumped mass model, where each floor is represented by an equivalent mass and stiffness. By applying seismic waves, the module estimates the response time history of each floor through a straightforward numerical calculation method. This approach simplifies the training process for the deep learning model in subsequent phases by avoiding direct fitting of complex data.
3. At the core of this framework lies the *Core Deep Learning Module*, designed to develop the Pretrained Model for predicting time history structural responses using as inputs structural dynamic features and seismic wave. Employing an autoregressive mechanism with a moving window during response prediction, the module reduces iteration error and expands the training set. This approach offers potential benefits in handling complex wave structures and potentially improves the overall prediction accuracy. During the training phase, data from numerical analysis is used to optimize the model parameters. However, during the inference phase, only seismic wave and structural dynamic features are required as inputs for rapid structural response prediction, which can be significantly faster than using FE analysis. This response predictive capability extends to various aspects such as displacement, acceleration, and inter-story drift. This information can then be used to assess potential damage to the building, which is critical for post-earthquake evaluation and informing necessary response actions.
4. Finally, to enhance accuracy further tailored to specific building types or regions, the Pretrained Model undergoes fine-tuning using LoRA (Low-Rank Adaptation of Large Language Models) to develop the *Finetuned Model*. This fine-tuning process with LoRA allows the model to adapt efficiently to new data while maintaining core knowledge learned during pre-training, potentially achieving superior accuracy.

Overall, the SeisGPT method combines physics-based structural dynamics with real-time response predictions, offering structural engineers an effective tool for mitigating seismic risks and designing more resilient buildings. Its real-time accuracy also streamlines the design process by enabling faster iterative analyses, improving efficiency, and reducing costs.

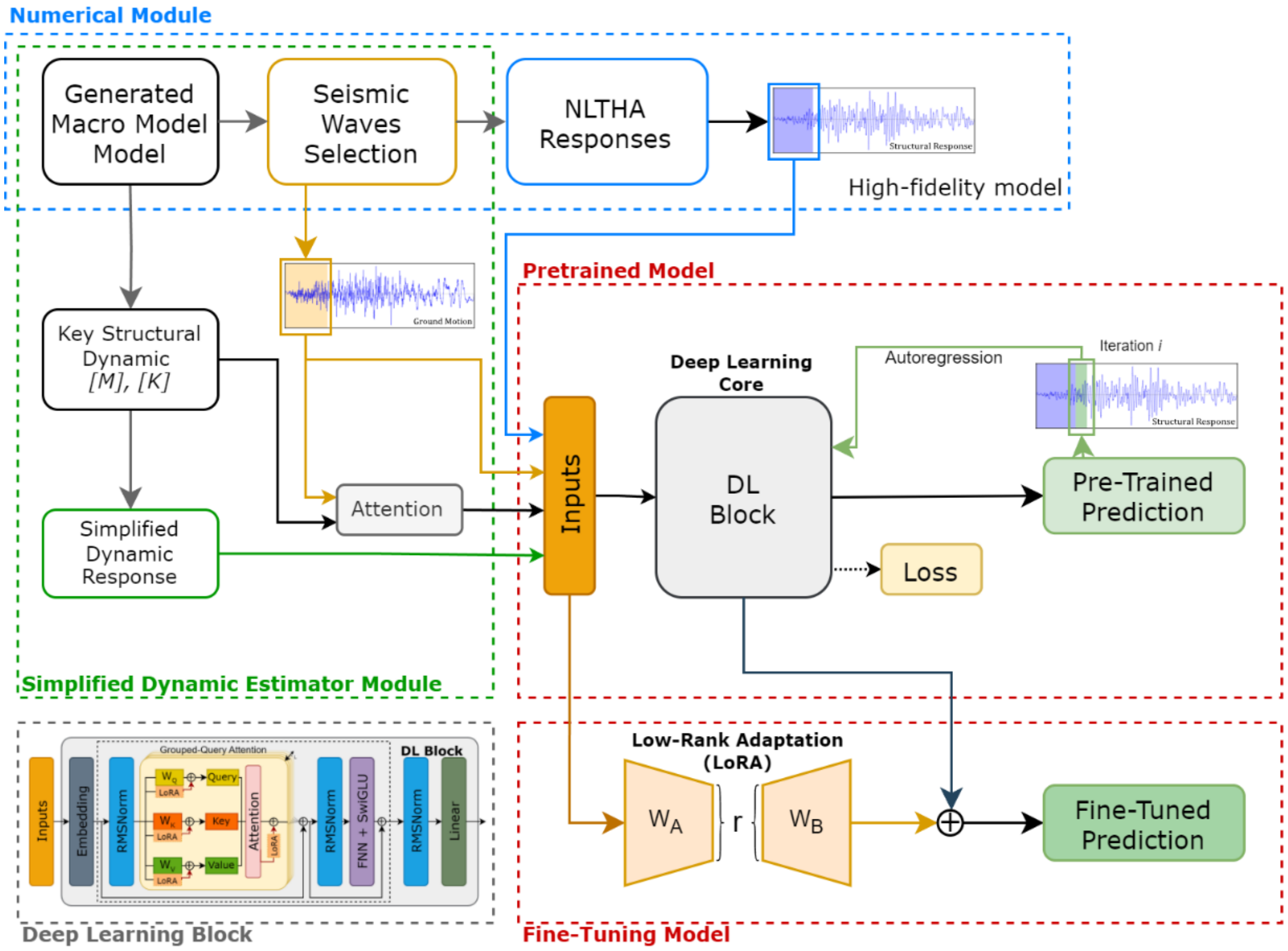

Figure 1 Schematic Overview of the SeisGPT Framework for Real-Time Seismic Response Prediction

## 2.2. Numerical Module: Dataset Acquisition and Processing

The development of a robust numerical module is crucial for creating a foundation for the deep learning model in this study. This module focuses on acquiring and processing high-quality data that accurately represents real-world seismic scenarios. Recognizing the critical role of high-fidelity data in enhancing deep learning performance, we developed a comprehensive dataset using Non-Linear Time History Analysis (NLTHA) simulations.

1- *Seismic Wave Selection*: To simulate realistic earthquake events, A total of 26,000 seismic waves selected from the Pacific Earthquake Engineering Research (PEER) Ground Motion Database. The distribution of these waves across different seismic intensities was designed to ensure a comprehensive representation of various seismic scenarios. Of the total waves, 48.1% (12,506 waves) corresponded to intensity 6, 41.77% (10,860 waves) to intensity 7, 8.86% (2,304 waves) to intensity 8, and 1.27% (330 waves) to intensity 9. This distribution aligns with recommended seismic design practices. The seismic wave inputs were sampled at a time interval of 0.02 seconds, providing high-resolution data for the subsequent NLTHA simulations.

2- *Building Models*: A total of 80,000 detailed macro models of buildings were created for this study. These models were classified into three main structural types: 50,000 frame structures, 20,000 frame-shear structures, and 10,000 shear wall complex structures. The building models represented a diverse range of typologies, including high-rise, mid-rise, and low-rise structures, and various functions such as office, residential, and commercial buildings.

Table 1 Structural Type Distribution

| Structure Type | Quantity | Number of Ground Motion | Direction | Number of Floors | Sub-total |
|---|---|---|---|---|---|
| Regular Frame Structure | 50,000 | 3 | 2 (x, y) | 1-10 | 300,000 |
| Regular Shear-Frame Structure | 20,000 | 3 | 2 (x, y) | 11-20 | 120,000 |
| Complex Shear-Frame Structure | 10,000 | 3 | 2 (x, y) | 20-33 | 60,000 |
| Total | 80,000 | **3** | 2 (x, y) | **-** | 480,000 |

The input variables for these building structures and the structural elements were defined according to the specifications outlined in GB/T50011-2010 [35] and JGJ 3-2010 [36]. To ensure a robust and versatile dataset, the range of values and combinations was intentionally extended beyond typical correlations, allowing for a more comprehensive exploration of the input space. This approach ensures that the model is trained on a diverse set of building configurations, enhancing its generalizability. The generation of each building geometries and configurations model adheres to a structured process that maintains realistic relationships between parameters while allowing for variability. Key geometric parameters include floor height (ranging from 3 to 3.6 m), slab thickness (80 to 120 mm), and the number of spans in the frame (from 3 to 10). The span length varies from 5 to 10 meters, with length-to-width ratios in the x and y directions ranging between 2/3 and 1. Shear wall thicknesses range from 200 to 400 mm. The dimensions of columns and beams are randomly generated in accordance with design codes, and calculations for axial compression and flexural strength are performed to ensure compliance. If any generated dimensions are deemed unreasonable, they are regenerated. The reinforcement ratio for all structural elements is automatically calculated according to the applicable design standards. The material parameters include concrete strength (C25 to C50) and reinforcement steel bar strength (355 to 400 MPa). This detailed generation process ensures that each building configuration was unique, realistic, and internally consistent, effectively representing real-world structural designs. The design parameters and their ranges for reinforced concrete (RC) buildings are summarized in Table 2.

Table 2 Design Parameters and Ranges for RC Buildings

| N | Parameter | Unit | Minimum | Maximum | Mean |
|---|---|---|---|---|---|
| ***Geometric and layout parameters*** | | | | | |
| 1 | Number of stories | - | 1 | 33 | - |
| 2 | Floor height | *m* | 3 | 3.6 | 2.9, 3.0, 3.1, 3.2, 3.3 |
| 3 | Slab thickness | *mm* | 80 | 150 | 90, 100, 110, 120 |
| 4 | Number of spans | - | 3 | 10 | 4, 5, 6, 7, 8, 9 |
| 5 | Length per span | *m* | 5 | 10 | 7.5 |
| 6 | Length-to-width ratio | - | 2/3 | 1 | 3/4, 4/5, 5/6 |
| 7 | Column size | *mm* | Randomly generated based on specifications, flexural strength calculated | | |
| 8 | Beam size | *mm* | Randomly generated based on specifications, flexural strength calculated | | |
| 9 | Shear wall length | *mm* | | | |
| 10 | Shear wall thickness | *mm* | 200 | 400 | 250, 300, 350 |
| 11 | Reinforcement | - | Automatically calculated according to specifications | | |
| ***Material parameters*** | | | | | |
| 12 | Concrete strength | *MPa* | C25 | C50 | C30, C35, C40, C45 |
| 13 | Rebar strength | *MPa* | 355 | 400 | 377.5 |

3- *NLTHA Simulations:* Each of the 80,000 building models was subjected to three randomly selected seismic waves in both the x and y directions, resulting in a total of 480,000 seismic response simulations. All simulations were carried out using OpenSees software, conducting NLTHA on all 480,000 individual scenarios to simulate their seismic responses. Implicit analysis was chosen for these simulations as it calculates the response at each time step assuming static equilibrium, making it well-suited for handling complex material behavior and large deformations, both of which are critical factors in earthquake simulations. The final output from the simulations is the NLTHA responses, which provide detailed seismic response data for each RC structure under the specified seismic waves. This response data is crucial for training, validating, and testing the deep learning models.

*4- Dataset Splitting and Characteristics:* To ensure robust model evaluation and generalizability, the dataset were divided into training, and testing sets using a random selection process. The final split allocated 90% (432,000 samples) for training data and 10% (48,000 samples) for testing data. Given the time interval of 0.02 seconds for each seismic wave and the duration of the simulations, the entire dataset represents approximately 15 billion data points. This vast and diverse dataset provides a solid foundation for model training and evaluation across different structural types and seismic scenarios.

The comprehensive nature of this dataset, encompassing various building types, seismic intensities, and structural responses, ensures that the resulting deep learning model will be well-equipped to handle a wide range of real-world scenarios. This robust numerical module forms the cornerstone of our study, enabling the development of a highly accurate and versatile seismic response prediction model.

## 2.3. Simplified Dynamic Response (SDR) Module

The primary focus of this study is the development and application of the Structural Dynamic Response (SDR) module, a pivotal tool [37] for estimating the dynamic responses of various structures. This involves representing the complex behavior of the frame with a more manageable model that retains essential characteristics. The module simplifies the complex building model into a lumped mass multi-degree-of-freedom (MDOF) system, where each floor is represented as a single mass point and the springs represent the stiffness of the structural elements. This approach reduces computational complexity while preserving the essential dynamic behavior of the structure. The SDRE module comprises two integral components: the Key Structural Dynamic Features and the Simplified Dynamic Response.

### 2.3.1. Key Structural Dynamic Features

First the Key Structural Dynamic Features play a foundational role by extracting key information regarding the structure's dynamics. This involves a comprehensive assessment of the mass and stiffness characteristics crucial for subsequent analysis. The determination of story mass encompasses the meticulous calculation of all structural elements' combined weight, including columns, walls, slabs, as well as any additional mass attributed to fixtures and live loads. Accurate estimation of story mass is paramount for ensuring the structural integrity and resilience of buildings, particularly in multi-story constructions where the distribution of mass between floors significantly impacts overall stability.

Similarly, the calculation of story stiffness is fundamental to understanding a structure's response to dynamic forces. This involves employing the original structure models, enabling a detailed analysis of stiffness characteristics of key components, such as beams, columns, and connections based on material properties and geometric configurations. Figure 2 illustrates a typical building structure, and the equivalent Multi-Degree-of-Freedom (MDOF) representation, highlighting the mass and stiffness properties.

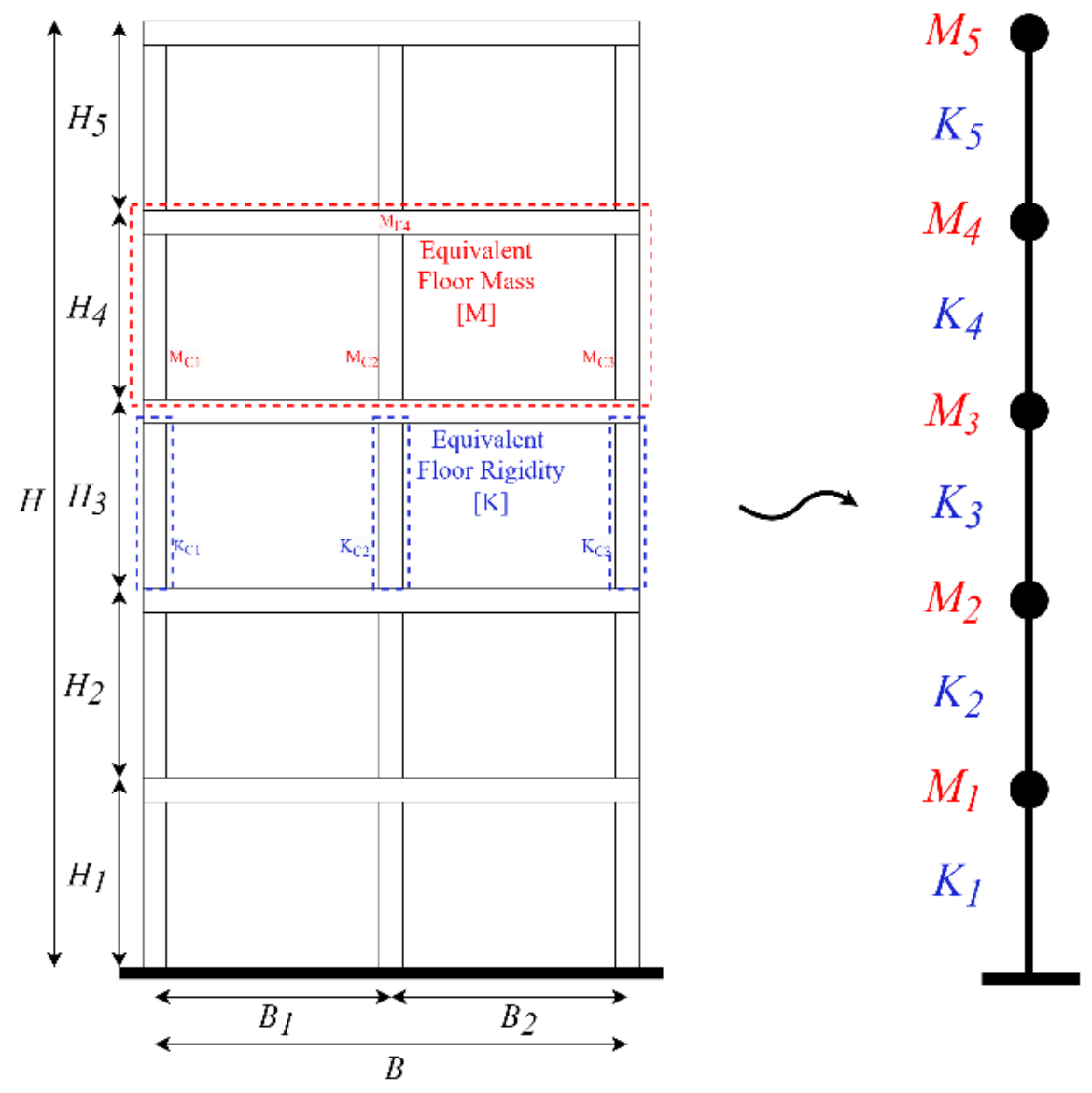

Figure 2 Building Structure Simplified to Multi-Degree-of-Freedom (MDOF)

Once the MDOF system is established, the next step is to identify the equivalent stiffness parameters. This is achieved through a parameter identification method that involves: Dynamic Response Analysis, Inverse Problem Formulation and Optimization Techniques. The dynamic response of the simplified MDOF system is analyzed under various loading conditions, including static and dynamic loads. The response is typically characterized by displacements, velocities, and accelerations at each mass. Then, the identification of stiffness parameters is framed as an inverse problem, where the goal is to determine the stiffness matrix that best matches the observed dynamic response of the original structure. This involves minimizing the difference between the calculated responses of the MDOF system and the actual responses of the original frame structure. Various optimization techniques, such as least squares or genetic algorithms, can be employed to solve the inverse problem. These techniques iteratively adjust the stiffness parameters until the best fit is achieved between the model predictions and the observed data.

To ensure the reliability of the identified stiffness parameters, the methodology incorporates a validation process where these parameters are compared with detailed finite element analysis results from the original structure. This step is essential for verifying that the simplified model accurately reflects the behavior of the actual structure, ensuring that its fundamental vibration period aligns with that of the real structure. If any discrepancies arise between the model's predictions and the observed data, the stiffness parameters are recalibrated. This iterative process refines the model, enhancing its accuracy and predictive capabilities.

Consequently, the periods of the simplified model and the FE model can be harmonized by scaling the overall stiffness of the simplified model using a coefficient S, calculated according to the following formula:

$$S = \left(\frac{T}{\acute{T}}\right)^2$$

where $S$ denotes the stiffness scaling factor, $T$ and $\acute{T}$ is the first period of the FE model and simplified model, respectively, calculated according to the following formula:

$$T = \frac{1}{f} = 2\pi\sqrt{\frac{M}{K}}$$

where $T$ indicates the main period of FE model in the selected direction, $M$ is the mass matrix, and $K$ is the stiffness matrix. This method is versatile and can be applied to a range of structural configurations, including different frame types such as moment-resisting frames, braced frames, and shear walls. It is capable of handling complex load distributions and varying mass distributions across different floors, which are common in real-world structures.

### 2.3.2.Simplified Dynamic Response

The Simplified Dynamic Response processes the reduced simplified model to estimate the nonlinear response of a structure's floor over time. Integrating story mass, stiffness, and structural dynamic equations, this module embeds physical parameters into the calculations, simplifying the problem of fitting deep learning models to structural responses under seismic activity.

The calculation method involves rapidly predicting floor response using a simplified numerical method, specifically the Newmark-β method. This method assumes linear acceleration changes within a time interval and incorporates floor displacement and acceleration expressed by the following equations:

$$u_{n+1} = u_n + \Delta t\dot{u}_n + \Delta t^2\left((1-2\beta)\frac{\ddot{u}_n}{2} + 2\beta\ddot{u}_{n+1}\right)$$

$$\dot{u}_{n+1} = \dot{u}_n + \Delta t\left((1-\gamma)\ddot{u}_n + \gamma\ddot{u}_{n+1}\right)$$

Here, $\gamma$ and $\beta$ are constants. The structural dynamics equation that must be satisfied is:

$$M\ddot{u}_{n+1} + C\dot{u}_{n+1} + Ku_{n+1} = G$$

By substituting equations (3) and (4) into equation (5), the equation for $\ddot{u}_{n+1}$ is obtained:

$$\ddot{u}_{n+1} = M^{-1}\left(G - C\left(\dot{u}_n + \Delta t(1-\gamma)\ddot{u}_n\right) - K\left(u_n + \Delta t\dot{u}_n + \Delta t^2(1-2\beta)\frac{\ddot{u}_n}{2}\right)\right)$$

Once $\ddot{u}_{n+1}$ is found, it is introduced back into equations (3) and (4) to calculate $u_{n+1}$

The damping matrix $C$ uses Rayleigh damping, expressed as:

$$C = \alpha M + \beta K$$

Parameters α and β are determined by:

$$\beta \ \alpha = 2\zeta\omega_1\omega_2 / (\omega_1 + \omega_2)$$

$$\beta = 2\zeta / (\omega_1 + \omega_2)$$

where $\zeta$ is the damping ratio, and $\omega_1$ and $\omega_2$ are the first- and second-order natural frequencies.

The structural dynamics equation is solved to obtain the nonlinear response, considering the damping matrix expressed via Rayleigh damping. This iterative process, based on the structural dynamic information extraction module, allows for accurate estimation of the nonlinear structural response.

### 2.4. SeisGPT Core Deep Learning Model

The core of SeisGPT is a sophisticated deep learning architecture specifically designed for processing sequential data, making it suitable for real-time earthquake response prediction. This core module is based on a unified Seismic-Response Feature Decoder (SRFD) architecture, utilizing a state-of-the-art decoder-only transformer model. The SRFD processes input response features and generates predictions while calculating the loss function for optimization during training.

The model takes and analyzes four inputs: earthquake wave data, history response, simplified model response, and dynamic characteristics represented by stiffness (*K*) and mass (*M*). The processing unfolds in two principal stages. In the initial stage, the model processes the earthquake wave data along with finite element (FE) time history responses and simplified model responses. In the second phase, the dynamic characteristics (stiffness *K* and mass *M*) are integrated into the earthquake wave data through an attention mechanism defined by the following equations:

$$attn_M = softmax\left(gelu\left(X \cdot W_q \cdot M \cdot (X \cdot W_k)^T\right)\right)$$

$$attn_K = softmax\left(gelu\left(X \cdot W_q' \cdot K \cdot (X \cdot W_k')^T\right)\right)$$

$$attn = attn_M + attn_K$$

$$X_{attn} = attn \cdot (X \cdot W_v)$$

where *X* represents the earthquake input data, $W_q$ and $W_q'$ is the query matrix for stiffness, $W_k$ and $W_k'$ is the key matrix for stiffness, *K* is the stiffness matrix, *Gelu* denotes the Gaussian Error Linear Unit activation function, and softmax refers to the *SoftMax* function. Predictions are generated, and the corresponding loss function is calculated based on the comparison between the predicted and actual data.

A defined equation aggregates the loss functions from all three blocks, creating a general loss function for the entire core module. This general loss function reflects the cumulative error across all analyses, which is backpropagated through the model to optimize it during training. Importantly, the output of this third stage serves as the final predicted results of the structural floor node responses.

Crucially, the training process leverages an autoregressive prediction strategy. In this approach, the model makes predictions in a step-by-step manner. It begins by receiving an initial segment of seismic wave data. Based on this data and the historical structural response, the model predicts the structural response for that segment. This predicted response is then fed back into the model along with the next segment of seismic wave data. The model iteratively builds on its previous predictions as it analyzes the entire seismic wave, using both the new seismic data and the previous response prediction for each subsequent segment. This approach enables SeisGPT to handle earthquakes of any length and potentially capture the evolving nature of the structural response during the event. This consistent and unified structure across all three blocks ensures efficient processing and integration of various types of input data for accurate real-time prediction of structural behaviors during seismic events.

*Seismic-Response Feature Decoder (SRFD)* is an advanced transformer-based neural network model. The model begins with an embedding layer that converts input tokens into higher-dimensional vector representations, increasing the number of channels to C=256 and timesteps to T=4096. This transformation allows the model to address data heterogeneity and handle diverse feature types, representing them in a format suitable for further processing. Root Mean Square (RMS) normalization is applied to these vectors to stabilize and normalize the inputs, particularly useful when dealing with time-series data.

$$\overline{a}_i = \frac{a_i}{RMS}, \text{ where } RMS = \sqrt{\frac{1}{d}\sum_{i=1}^{d} a_i^2}$$

where $\overline{a}_i$ Normalized Activation Value, $a_i$ Original Activation Value, and d Dimension of Token Vector.

A key feature of the model is the Grouped Query Attention (GQA) mechanism, a variation of the standard multi-head attention. GQA segments queries into groups that share a single set of keys and values heads, enhancing both performance and memory efficiency. This mechanism plays a crucial role in focusing on relevant features by grouping input features and using self-attention to identify the most important elements within each group. This prioritization of informative features significantly contributes to the overall model performance. Rotary Position Embedding (RoPE) is applied to queries and keys within the GQA mechanism to effectively encode positional information. The output from this attention layer then passes through a feed-forward network (FNN) utilizing the SwiGLU (Sigmoid-Weighted Linear Unit) as activation function, which has been shown to be more effective than the traditional ReLU activation function in certain cases.

$$SwiGLU(x) = x \cdot \sigma(x) \text{ where } \sigma(x) = \frac{1}{1+e^{-x}}$$

Residual connections and additional RMS normalization are employed after both the attention and feed-forward layers to aid gradient flow and model training. This block configuration is repeated N times, indicating the stacking of transformer blocks to deepen the model for learning complex patterns. Post these repetitions, the output undergoes another RMS normalization, followed by a linear transformation to map it to the desired dimensions. This architecture showcases the integration of modern techniques like Grouped Query Attention and SwiGLU to enhance performance and efficiency in processing sequential data.
The SRFD's sophisticated design allows it to effectively manage diverse and complex inputs, making it a powerful tool for real-time earthquake response prediction.

## 3. Training Strategy & Evaluation Metrics

The SeisGPT method offers a data-driven approach that integrates physics-based structural dynamic features with real-time response prediction, making it a valuable tool for structural engineers in mitigating seismic risks and designing buildings with enhanced resilience.

Response prediction is achieved through an autoregressive iterative prediction mechanism. Instead of processing the entire seismic wave directly, the approach involves segment-wise iterative processing of the input seismic wave. During each iteration, the predicted structural response output serves as an input component for the subsequent prediction. This segment-wise processing simplifies the model's training and enables it to predict structural response time histories for seismic wave inputs of any length.

By fixing the length of the single input and output data in the time dimension, the model's training process is streamlined. The model captures temporal dependencies in the data by predicting the time history of structural responses one time step at a time. Advanced optimization algorithms such as Adam and RMSprop are utilized to dynamically adjust the learning rate, ensuring efficient parameter updates and faster convergence. These algorithms help the model achieve better performance by fine-tuning the training process.

The performance of the models in regression tasks was evaluated using several performance metrics. The training employs a custom loss function that combines mean squared error (MSE). This loss function ensures that the model not only fits the training data well but also generalizes to unseen data by penalizing overfitting. By incorporating these three blocks, the loss function ensures that the model not only fits the data points accurately but also captures the temporal dynamics and changes in the data. This comprehensive approach leads to a more robust and reliable predictive model, capable of providing precise time history predictions for structural responses under seismic loading conditions. The optimization of this loss function during training adjusts the model parameters to minimize these errors, resulting in high fidelity and accuracy in the model's predictions.

However, MSE alone is not always sufficient to assess the performance of neural networks on data, as it is sensitive to outliers and not expressed in the same units as the target variable. Therefore, mean absolute error (MAE), Mean Relative Error (MRE) and Pearson correlation coefficient (R), are also used as metrics to measure different aspects of the errors. These metrics are calculated using the following formulas:

$$MSE = \frac{1}{n}\sum_{i=1}^{n}(T_i - O_i)^2$$

$$MRE = \frac{1}{n}\sum_{i=1}^{n}\frac{|T_i - O_i|}{|T_i|}$$

$$MAE = \frac{1}{n}\sum_{i=1}^{n}|T_i - O_i|$$

$$R = \frac{\sum_{i=1}^{n}(T_i - \bar{T}_i)(O_i - \bar{O}_i)}{\sqrt{\sum_{i=1}^{n}(T_i - \bar{T}_i)^2 \sum_{i=1}^{n}(O_i - \bar{O}_i)^2}}$$

Here, *n* is the number of data points (number of time-steps); Ti and Oi are the target and the predicted value, respectively. Ti and Oi are the mean of the target and mean of the predicted values, respectively. Lower values of MSE, MAE, indicate a better model fit to the data. Higher R values (closer to 1) suggest a stronger model fit, capturing more of the data's variability. Lower R values imply a weaker fit, explaining less of the data variance.

## 4. Results & Discussions

The overall performance of the SeisGPT models, in their pretrained state, was rigorously evaluated using an unseen test dataset comprising 48,000 samples. These samples, distinct from the training data, included various building structures subjected to seismic ground motion excitation, providing a diverse set of challenges for model evaluation. This approach enabled the assessment of the model's generalizability across different structural typologies and material properties.

Key performance metrics such as MSE, MRE, and R were used to evaluate the model's predictive accuracy. These metrics offer a comprehensive understanding of the model's ability to predict both displacement and acceleration responses.

For displacement prediction, the SeisGPT-Base model achieved an MSE of 0.0105, an MRE of 0.0576, and an R value of 0.926, demonstrating strong predictive performance. In acceleration prediction, the model recorded an MSE of 0.0025, an MRE of 0.0259, and an R value of 0.949, highlighting its robustness in accurately forecasting seismic responses. The high correlation between predicted and actual values reinforces the model's reliability and potential for practical earthquake engineering applications, providing accurate predictions across diverse building typologies and seismic scenarios.

The detailed performance results of the pretrained model are summarized in Table 3, with further case-specific results available in the following section.

Table 3 Performance metrics across the entire dataset

| **Model** | **Performance** | **Acceleration** | | | **Displacement** | | |
|---|---|---|---|---|---|---|---|
| | | **MSE** | **MRE** | **R** | **MSE** | **MAE** | **R** |
| **Pretrained** | Overall | 0.0025 | 0.0259 | 0.949 | 0.0105 | 0.0576 | 0.926 |

## 5. Case Study Descriptions

To further evaluate the performance of the SeisGPT-Base model, rigorous evaluations were conducted on three distinct building types: a regular frame structure, a regular shear-frame structure, and a complex shear-frame structure. Each building type was subjected to one or two unique seismic waves in two direction X and Y, thereby constituting six unseen test cases. Comprehensive details regarding their material properties, number of stories, and other relevant structural aspects are provided in the supplementary material.

### 5.1 Numerical Example 1: Frame Structure

The first case study involved a 4-story, low-rise residential building with a RC frame structure, standing approximately 12 meters high. The building's displacement response was analyzed under two ground motion accelerations (GMAs), RSN14985 and RSN15907, in both the x and y directions.

The SeisGPT model accurately predicted the displacement responses at various floors under these ground motions, showing a strong correlation between predicted and actual displacements. Figure 3 and Figure 4 present the displacement plots for the middle (2$^{nd}$ floor) and top floors (4$^{th}$ floor) in both directions, demonstrating close agreement between the predictions and actual responses.

For GMA 1 (RSN14985), the model achieved MSE values of 0.019 (x direction) and 0.036 (y direction), with MAE values of 0.092 and 0.123, respectively. The correlation coefficients (R) were 0.987 for both directions. For GMA 2 (RSN15907), the MSE values were 0.048 (x direction) and 0.049 (y direction), and the MAE values were 0.128 and 0.134, respectively. The correlation coefficients remained consistent at 0.987 and 0.986 for the x and y directions. These results validate the model's ability to accurately predict the dynamic response of low-rise buildings under seismic conditions.

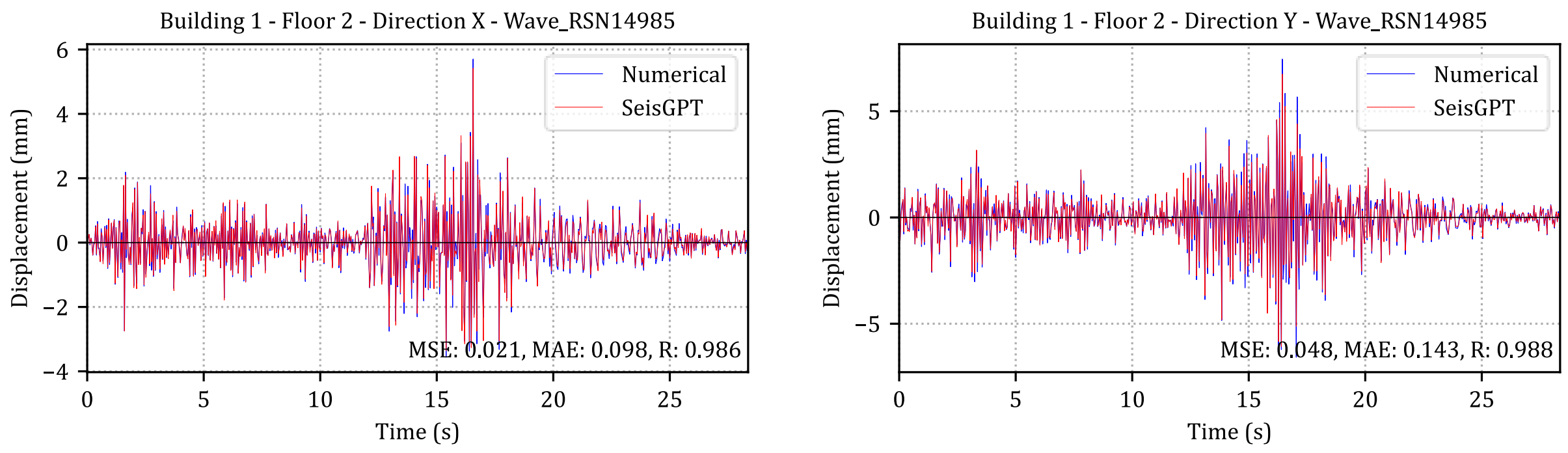

Direction X Direction Y

a) Middle Floor

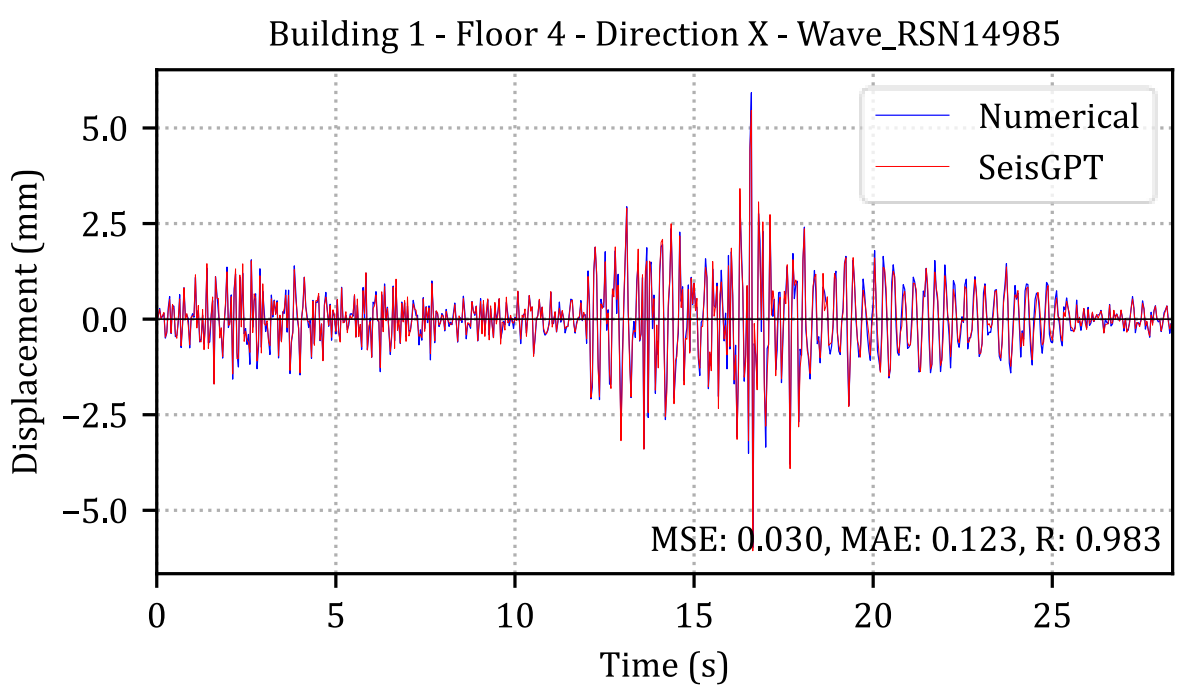

Direction X

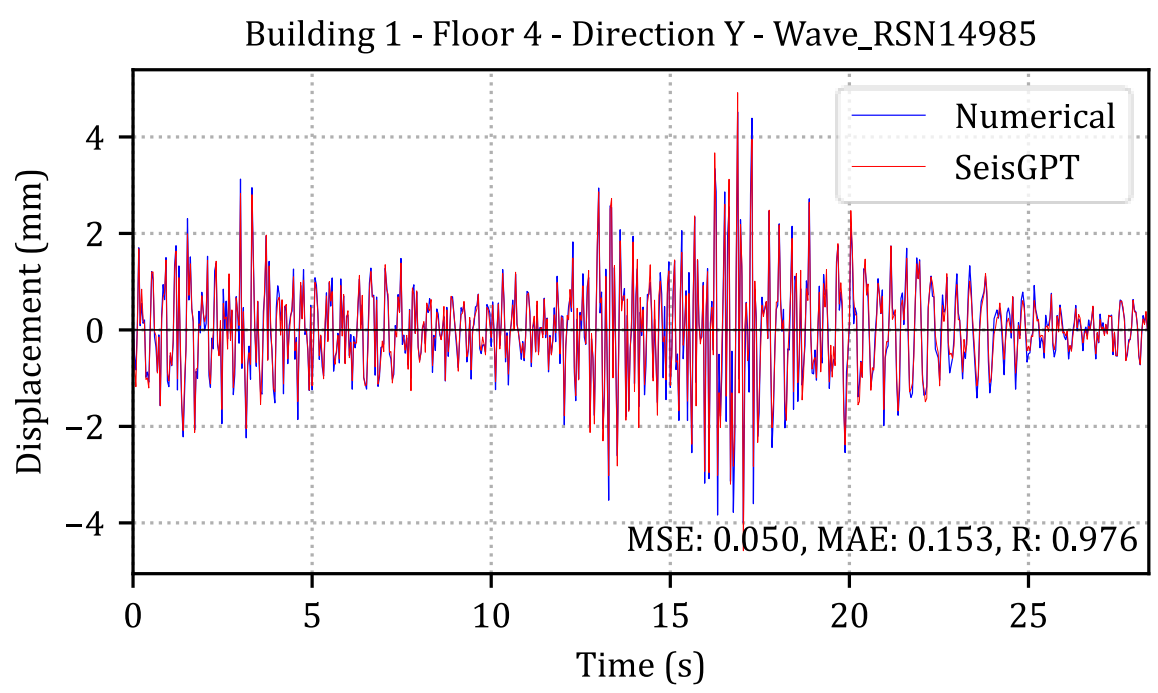

Direction Y

b) Top Floor

Figure 3 Comparison of Structural Responses (Case 1 & Wave_RSN14985)

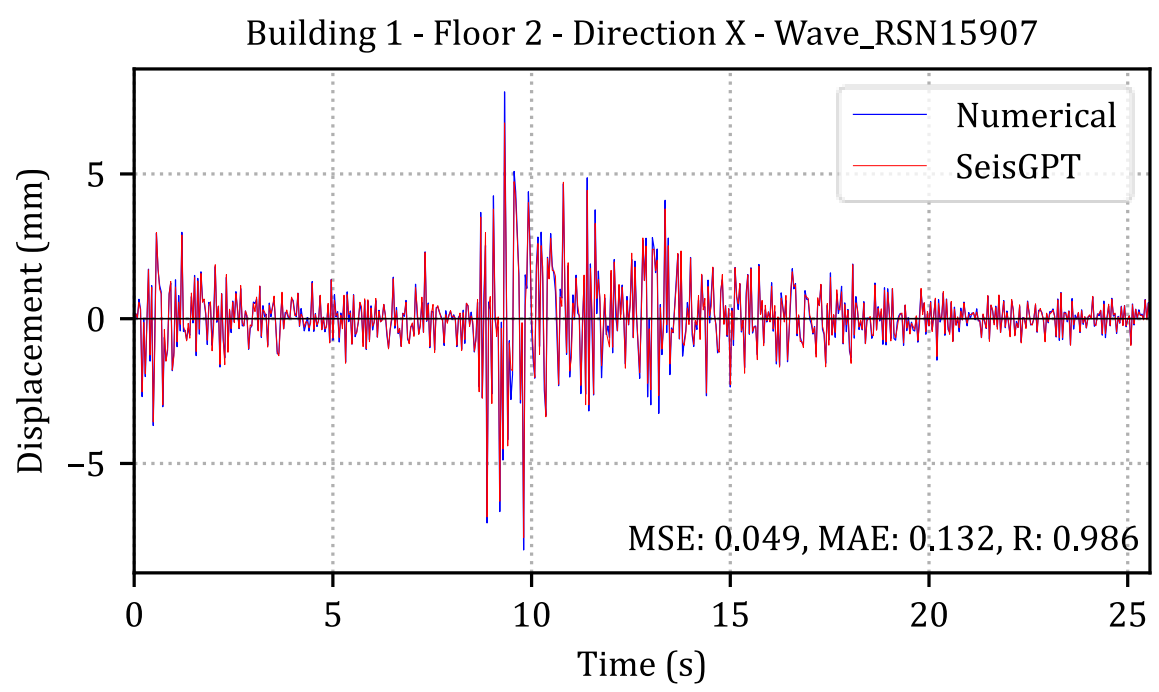

Direction X

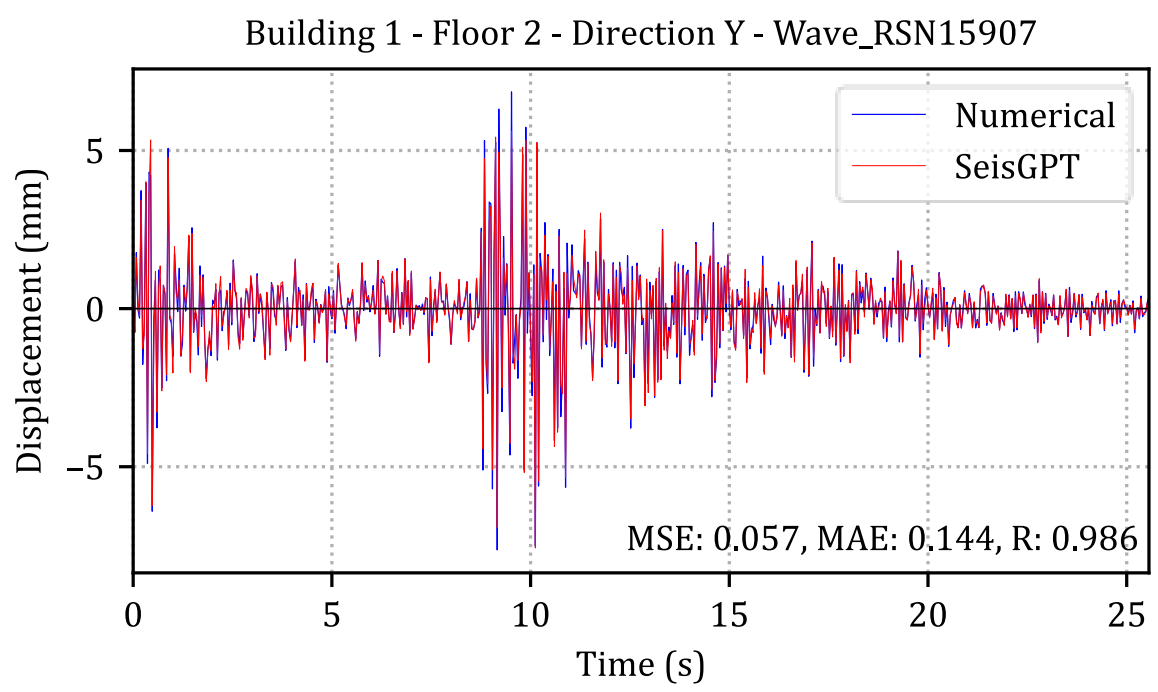

Direction Y

a) Middle Floor

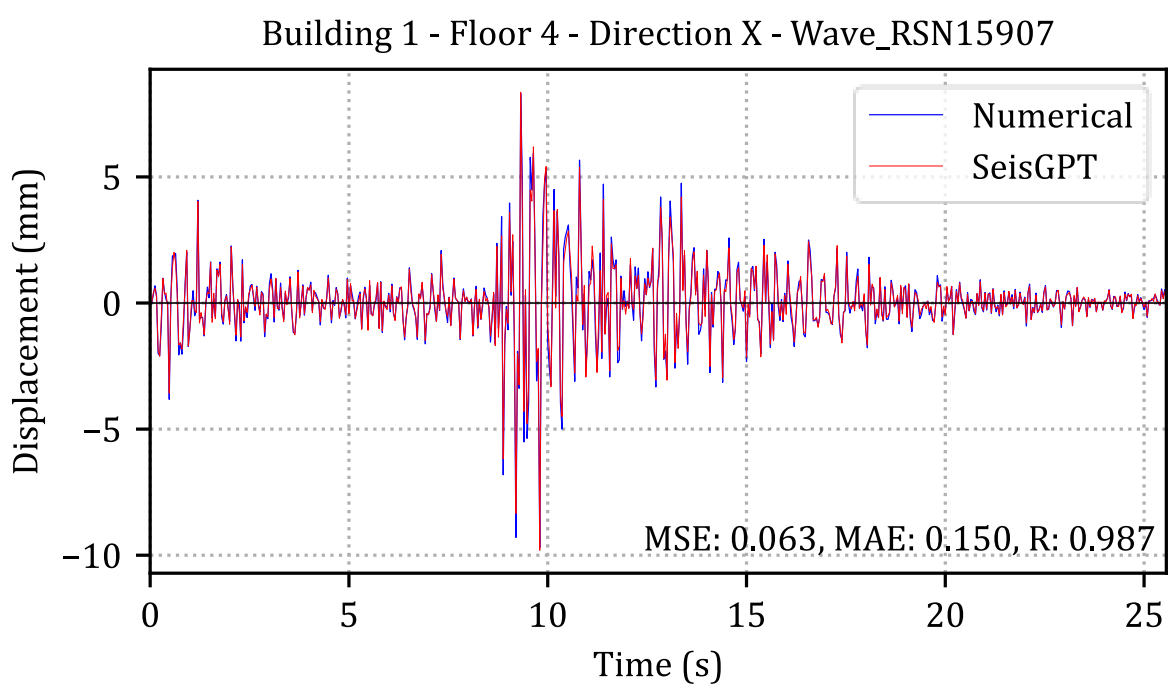

Direction X

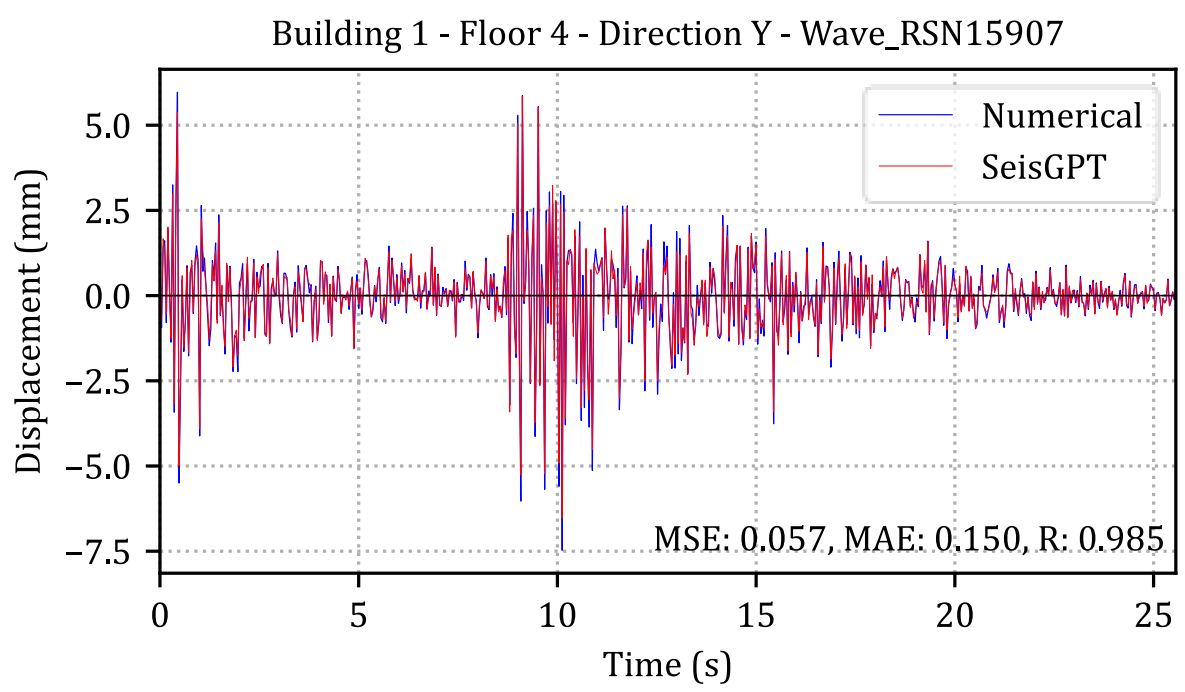

Direction Y

b) Top Floor

Figure 4 Comparison of Structural Responses (Case 1 & Wave_ RSN15907)

## 5.2 Numerical Example 2: Shear-Frame Structure

The second case study examined an 11-story, mid-rise RC shear wall building, measuring 33 meters in height. The building was subjected to a single ground motion acceleration (GMA) RSN11727 in both the x and y directions. As shown in Figure 5, the predicted displacement responses at the 5th and 11th floors closely matched the actual results.

For this case, the model produced MSE values of 0.030 (x direction) and 0.015 (y direction), with MAE values of 0.104 and 0.067, respectively. The correlation coefficients (R) were 0.979 for the x direction and 0.977 for the y direction. These findings affirm the model's effectiveness in predicting the behavior of mid-rise RC shear-wall buildings under seismic loading.

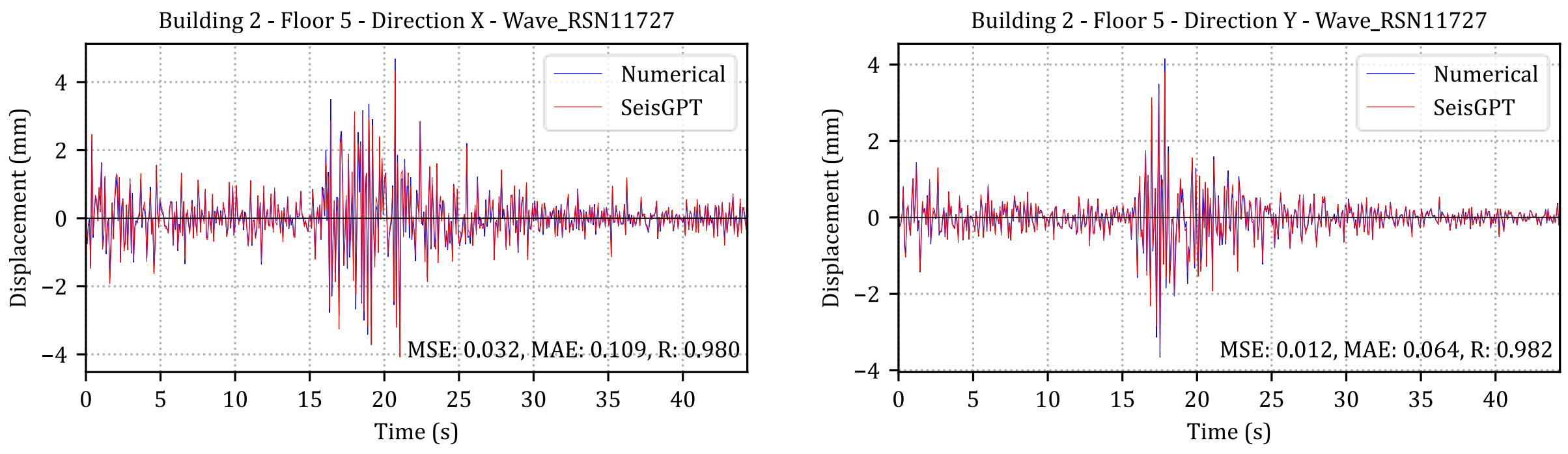

Direction X Direction Y

a) Middle Floor

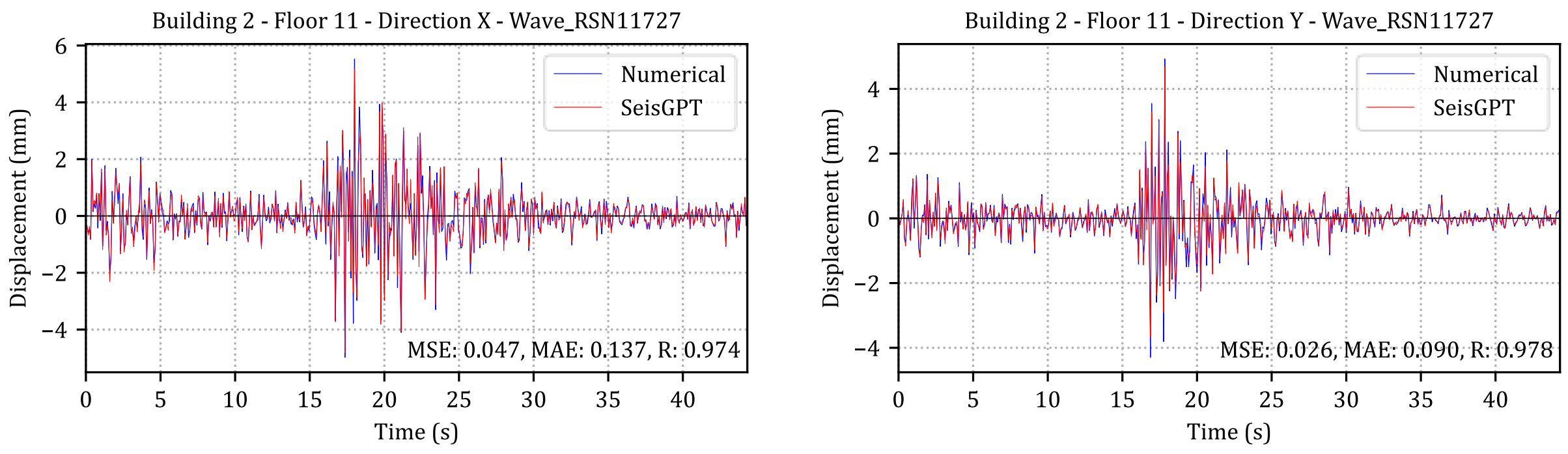

Direction X Direction Y

b) Top Floor

Figure 5 Comparison of Structural Responses (Case 2 & Wave_RSN11727)

### 5.3 Numerical Example 3: Complex Shear Wall Structure

The third case involved a 12-story RC shear wall structure, approximately 36 meters tall, subjected to ground motion acceleration RSN04234 in both the x and y directions. The SeisGPT model once again demonstrated strong predictive accuracy, with high correlations observed between the predicted and actual displacement responses across various floors.

As illustrated in Figure 6, the time series plots show close agreement between SeisGPT predictions and finite element analysis (FEA) results for both the middle and top floors. The model achieved MSE values of 0.018 (x direction) and 0.007 (y direction), with MAE values of 0.083 and 0.056, respectively. The correlation coefficients (R) were 0.953 (x direction) and 0.954 (y direction). These results highlight the SeisGPT model's strong potential for accurately predicting seismic responses in complex structures, offering significant practical applications in earthquake engineering.

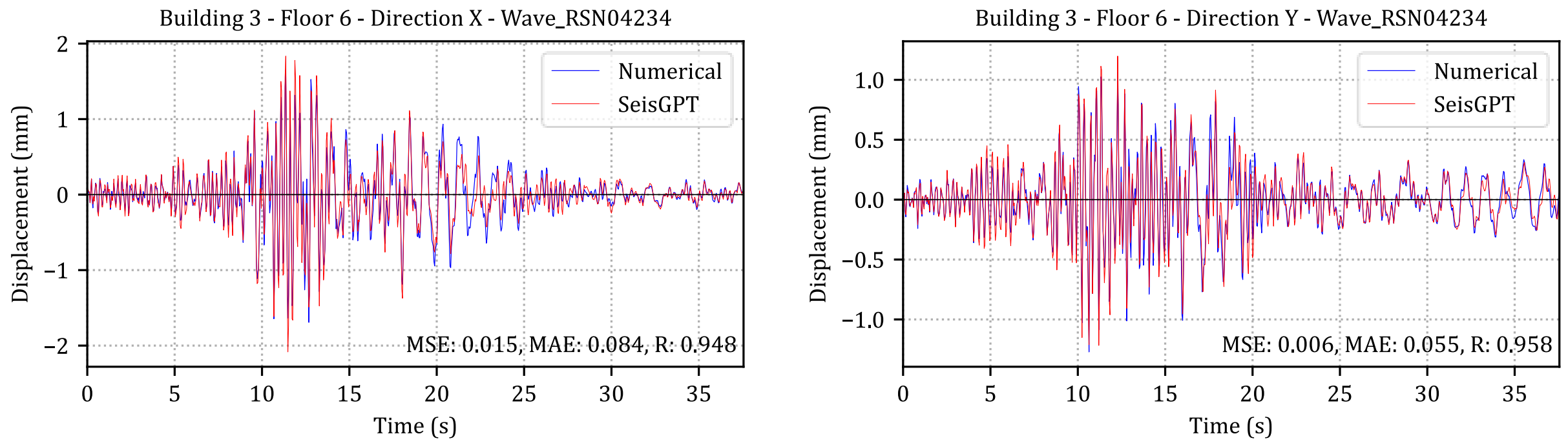

Direction X Direction Y

a) Middle Floor

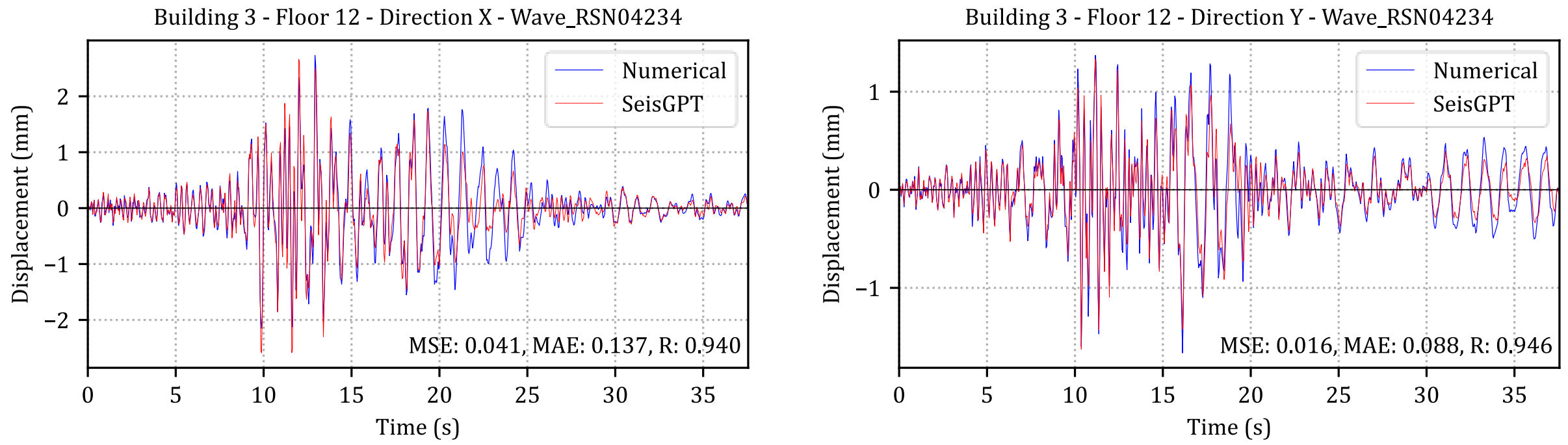

Direction X Direction Y

b) Top Floor

Figure 6 Comparison of Structural Responses (Case 3 & Wave_RSN04234)

Overall, the SeisGPT-Base model has proven highly effective in predicting the seismic displacement responses of different structural types, from low-rise frame structures to more complex shear-wall designs. Across all test cases, the model consistently demonstrated strong correlations with actual structural responses and produced low error metrics, including MSE and MAE. This suggests that SeisGPT can reliably predict the dynamic behavior of various building types under seismic loading, making it a promising tool for earthquake engineering and structural analysis. The results underscore the model’s robustness and versatility, showing potential for practical use in structural design and performance evaluation in earthquake-prone regions.

## 6. Real-Time Prediction and Computational efficiency of the proposed approach

The SeisGPT-Base model demonstrated exceptional accuracy in predicting structural responses in real-time under diverse loading conditions and across all floors within a structure. Its strong generalization capabilities to unseen data highlight its potential for practical applications in structural health monitoring and earthquake engineering. By combining high precision, real-time predictions, and remarkable computational efficiency, SeisGPT stands out as a valuable tool for seismic response analysis across a wide range of building configurations. The SeisGPT-Base model was trained on a powerful GPU cluster consisting of four machines, each equipped with eight H800 GPUs. This robust hardware setup enabled the model to deliver real-time predictions while achieving exceptional computational efficiency. In contrast to traditional finite element analysis (FEA), which can take several hours to produce results, SeisGPT provides instant predictions with comparable accuracy. This significant improvement in computational efficiency drastically reduces the time and resources required for seismic response assessments.

The pretrained SeisGPT model can predict response time histories for various building typologies and materials in real time, making it particularly valuable for structural health monitoring and earthquake engineering, where timely and accurate predictions are critical.

## 7. Conclusion

This study introduces SeisGPT, an innovative framework leveraging deep learning to predict the dynamic responses of building structures under seismic forces. By utilizing the Generative Pre-trained Transformer (GPT) architecture, SeisGPT offers a significant advancement in balancing prediction accuracy and computational efficiency, addressing the limitations of traditional finite element methods (FEM).

SeisGPT integrates physics-based structural dynamic features with a data-driven approach to enhance seismic response prediction accuracy. The model training process involves acquiring and processing a comprehensive dataset of 26,000 seismic waves from the PEER Ground Motion Database, which are partitioned into training, validation, and testing sets. These seismic waves are used to simulate realistic earthquake events using high-fidelity macro models of various building typologies, analyzed with OpenSees software. The resulting NLTHA responses provide detailed seismic response data crucial for model training.

The Structural Dynamic Response (SDRE) module simplifies complex macro models into a lumped mass multi-degree-of-freedom (MDOF) system, reducing computational complexity while preserving essential dynamic behavior. Key structural dynamic features, such as story mass and stiffness, are extracted for subsequent analysis. This simplified approach enables the SDRE module to accurately predict the dynamic responses of structures subjected to seismic excitations.

The pretrained SeisGPT model demonstrated strong predictive performance across both acceleration and displacement metrics. For acceleration, it achieved a Mean Squared Error (MSE) of 0.0025, a Mean Relative Error (MRE) of 0.0259, and a correlation coefficient (R) of 0.949. For displacement, the model produced an MSE of 0.0105, a Mean Absolute Error (MAE) of 0.0576, and an R value of 0.926. These metrics reflect the model's high precision and reliability when applied to new data, confirming its suitability for practical earthquake engineering applications.

Comparative studies demonstrated SeisGPT's superior performance over other competitive baselines, particularly in terms of precision, computational efficiency, and applicability across diverse building typologies and seismic intensities. The rigorous validation on real-world scenarios showcased the framework's robustness and reliability in accurately predicting structural responses under seismic actions.

One limitation of the SeisGPT model is its dependency on high-quality, diverse training data to generalize well across different building typologies and seismic intensities. Future work should focus on expanding the dataset to include more diverse seismic scenarios and building structures.

## Acknowledgments

The authors gratefully acknowledge the financial support from the National Key Research and Development Program of China (Grant No. 2023YFC3805000), the Distinguished Young Scientist Fund of National Natural Science Foundation of China (Grant No.52025083), the XPLORE PRIZE (Grant No. XP202342) , and the Shanghai Urban Digital Transformation Special Fund (Grant No. 202201033).